\documentstyle[sprocl]{article}
\input epsf.tex

\def\sst{\scriptscriptstyle}
\def\pref#1{(\ref{#1})}

\def\eps{\epsilon}

\def\ie{{\it i.e.}}

\def\hf{\frac12}
\def\nth#1{\frac{1}{#1}}

\def\be{\begin{equation}}
\def\ee{\end{equation}}
\def\bea{\begin{eqnarray}}
\def\eea{\end{eqnarray}}
\begin{document}
\rightline{hep-ph/9812470}
\rightline{McGill-98/39}
\vspace{10mm}
\title{An Ode to Effective Lagrangians}
\author{C.P. Burgess}
\address{Physics Department, McGill University\\
3600 University Street, Montr\'eal, Qu\'ebec, Canada, H3A 2T8\\
E-mail: cliff@physics.mcgill.ca} 


\maketitle\abstracts{A brief introduction is given to
the methods and spirit of effective lagrangians. The emphasis
is on a summary of the overall picture, using a simple model
as the vehicle to motivate and illustrate the main points. 
Powercounting is illustrated by estimating the size of 
the quantum corrections to the predictions of classical gravity.  
(Invited talk presented to the conference Radcor 98, Barcelona, 
September 1998.)}

\section{Introduction and Summary}

In all branches of theoretical physics a key part of any good
prediction is a careful assessment of the theoretical error which
the prediction carries. Such an assessment is a precondition for any
detailed quantitative comparison with experiment. As is 
clear from the much of the work presented at this meeting, in mature
theories like the Standard Model this assessment of error usually can
be reliably determined based on an understanding of
the small quantities which control the validity of the approximations
used when making predictions. 

\subsection{`Unreasonably' Good Predictions}

It sometimes happens that predictions are much more accurate
than would be expected based on an assessment of 
the approximations on which they appear to be based. A
famous example of this is encountered in the precision tests of
Quantum Electrodynamics, where the value of the
fine-structure constant, $\alpha$, was until recently, 
obtained using the Josephson effect in superconductivity.

A DC potential difference applied at the boundary between 
two superconductors can produce an AC Josephson current whose
frequency is precisely related to the size of the applied
potential and the electron's charge. Precision measurements
of frequency and voltage are in this way converted into
a precise measurement of $e/\hbar$, and so of $\alpha$.
But use of this effect to determine $\alpha$ 
only makes sense if the predicted relationship between
frequency and voltage is also known to an
accuracy which is better than the uncertainty in $\alpha$. 

It is, at first sight, puzzling how such an accurate prediction
for this effect can be possible. After all, the prediction
is made within the BCS theory of superconductivity, \cite{BCS} which
ignores most of the mutual interactions of electrons, focussing
instead on a particular pairing interaction due to phonon 
exchange. Radical though this approximation might appear
to be, the theory works rather well (in fact, surprisingly
well), with its predictions often agreeing with experiment to 
within several percent. But expecting successful predictions with an
accuracy of parts per million or better would appear to be
optimistic indeed!

\subsection{The Low-Energy Approximation}

The astounding accuracy required to successfully predict the
Josephson frequency may be understood at another level, 
however. The key observation is that this prediction does
not rely at all on the details of the BCS theory, depending
instead only on the symmetry-breaking pattern which it
predicts. Once it is known that a superconductor 
spontaneously breaks the $U(1)$ gauge symmetry of electromagnetism,
the Josephson prediction follows on general grounds in the
low-energy limit.\cite{WeinbergSC} The validity of the prediction is therefore
not controlled by the approximations made in the BCS theory,
since {\it any} theory with the same low-energy symmetry-breaking
pattern shares the same predictions. 

The accuracy of the predictions for the Josephson effect
are therefore founded on symmetry arguments, and on
the validity of a low-energy approximation. Quantitatively,
the low-energy approximation involves the neglect of powers of the
ratio of two scales, $\omega/\Omega$, where $\omega$ is
the low energy scale of the observable under 
consideration --- like the applied voltage in the 
Josephson effect --- and $\Omega$ is the higher energy 
scale --- such as the superconducting gap 
energy --- which is intrinsic to the system under study.

Indeed, arguments based on a similar
low-energy approximation may also be used to explain the
surprising accuracy of many other successful models throughout
physics, including the BCS theory itself.\cite{Polchinski,Shankar,Frohlich} 
This is accomplished by showing that only the specific 
interactions used by the BCS theory are relevant at low energies, 
with all others being suppressed in their effects by powers of a 
small energy ratio. 

Although many of these arguments were undoubtedly known in various
forms by the experts in various fields since very early days, 
the systematic development of these arguments into precision
calculational techniques has happened more recently. With this
development has come considerable cross-fertilization of 
techniques between disciplines, with the realization that the same 
methods play a role across diverse disciplines within physics. 

The remainder of this lecture briefly summarizes the
techniques which have been developed to exploit low-energy
approximations. These are most efficiently expressed
using effective-lagrangian methods, which are designed
to take advantage of the simplicity of the low-energy limit
as early as possible within a calculation. The gain in
simplicity so obtained can be the decisive difference
between a calculation's being feasible rather than 
being too difficult to entertain. 

Besides providing this kind of practical advantage, 
effective-lagrangian techniques also bring real conceptual 
benefits because of the clear separation they permit between
of the effects of different scales. Both of these kinds of
advantages are illustrated here using explicit examples. First
\S2\ presents a toy model involving two spinless particles to
illustrate the general method, as well as some of its 
calculational advantages. This is followed by a short 
discussion of the conceptual advantages, with 
quantum corrections to classical general relativity, 
and the associated problem of the nonrenormalizability 
of gravity, taken as the illustrative example. 

\section{A Toy Example}

In order to make the discussion as concrete as possible,
consider the following model for a single complex scalar field, $\phi$:
\begin{eqnarray}
\label{abeltoymodel}
{\cal L} &=& - \partial_\mu \phi^* \partial^\mu \phi - 
V(\phi^* \phi), \nonumber\\
\hbox{with} \qquad V &=& {\lambda^2 \over 4} \; 
\left( \phi^* \phi - v^2 \right)^2 .
\end{eqnarray}
This theory enjoys a continuous $U(1)$ symmetry of the form 
$\phi \to e^{i\omega} \; \phi$, where the parameter, $\omega$, 
is a constant. The two parameters of the model are 
$\lambda$ and $v$. Since $v$ is the only dimensionful 
quantity it sets the model's overall energy scale.

The semiclassical approximation is justified if the dimensionless 
quantity $\lambda$ should be sufficiently small. In this approximation
the vacuum field configuration is found by minimizing the system's 
energy density, and so is given (up to a $U(1)$ transformation)
by $\phi = v$. 
For small $\lambda$ the spectrum consists of two weakly-interacting
particle types described by the fields ${\cal R}$ and ${\cal I}$,
where $\phi = \left( v + \nth{\sqrt2} \; {\cal R} \right) + 
\frac{i}{\sqrt2} \; {\cal I}$. To leading order in $\lambda$
the particle masses are $m_{\sst I} = 0$ and $m_{\sst R} 
= \lambda v$. 

The low-energy regime in this model is $E \ll m_{\sst R}$. The
masslessness of ${\cal I}$ ensures the existence of degrees of
freedom in this regime, with the potential for nontrivial
low-energy interactions, which we next explore. 

\subsection{{\cal I} -- {\cal I} Scattering}

The interactions amongst the particles in this model 
are given by the scalar potential :
\begin{equation}
\label{potl}
V = {\lambda^2 \over 16} \; \Bigl( 2 \sqrt2 \, v {\cal R} +
{\cal R}^2 + {\cal I}^2 \Bigr)^2. 
\end{equation}

Imagine using the potential of eq.~\pref{potl} to 
calculate the amplitude for
${\cal I}$--${\cal I}$ scattering at low energies 
to lowest-order in $\lambda$. 
The $S$-matrix obtained by evaluating the four tree-level diagrams
is proportional to the following invariant amplitude:
\begin{eqnarray}
\label{smatrixresult}
{\cal A} &=& - \; {3 \lambda^2 \over 2} + \left( { \lambda^2 v \over \sqrt2}
\right)^2 \left[ {1 \over (s+r)^2 + m_{\sst R}^2 - i\eps}  \right. \nonumber\\
&& \qquad\qquad \left. + {1 \over (r-r')^2 + m_{\sst R}^2 -i \eps} + 
{1 \over (r - s')^2 + m_{\sst R}^2 - i\eps}\right], 
\end{eqnarray}
where $s^\mu$ and $r^\mu$ (and $s'{}^\mu$ and $r'{}^\mu$)
are the 4-momenta of the initial (and final) particles.

An interesting feature of this amplitude is that when it is 
expanded in powers of four-momenta, both its leading and
next-to-leading terms vanish. That is: 
\begin{eqnarray}
\label{zeromomlim}
{\cal A} &=&  \left[ - \; {3 \lambda^2 \over 2} + {3 \over
m_{\sst R}^2} \, \left( { \lambda^2 v \over \sqrt2}
\right)^2 \right] + {2 \over m_{\sst R}^4} \;
\left( { \lambda^2 v \over \sqrt2} \right)^2 \Bigl[- r \cdot s
+ r \cdot r' + r \cdot s' \Bigr]  \nonumber\\
&& \qquad\qquad\qquad\qquad \qquad\qquad\qquad
+ O(\hbox{quartic in momenta}) \nonumber\\
&=& 0 +O(\hbox{quartic in momenta}) .
\end{eqnarray}
The last equality uses conservation of 4-momentum: $s^\mu + r^\mu
= s'{}^\mu + r'{}^\mu$ and the massless mass-shell condition
$r^2 = 0$. 

Clearly the low-energy particles interact more weakly than
would be expected given a cursory inspection of the scalar
potential, eq.~\pref{potl}, since at tree level 
the low-energy scattering rate is suppressed by 
at least eight powers of the small energy ratio $r = E/m_{\sst R}$. 
The real size of the scattering rate might depend crucially
on the relative size of $r$ and $\lambda^2$, should the vanishing
of the leading low-energy terms turn out to be an artifact
of leading-order perturbation theory. 

If ${\cal I}$ scattering were of direct experimental interest, 
one can imagine considerable effort being invested in obtaining
higher-order corrections to this low-energy result. And the
final result proves to be quite interesting: as may be verified by 
explicit calculation, the first two terms in the low-energy
expansion of ${\cal A}$ vanish order-by-order in perturbation theory. 
Furthermore, a similar suppression turns out also to hold 
for all other amplitudes involving ${\cal I}$ particles, with
the $n$-point amplitude for ${\cal I}$ scattering being
suppressed by $n$ powers of $r$. 

Clearly the hard way to understand these low-energy
results is to first compute to all orders in $\lambda$ and 
then expand the result in powers of $r$. A much more
efficient approach exploits the simplicity of small $r$
{\sl before} calculating scattering amplitudes. 

\section{The Toy Model Revisited}

The key to understanding this model's low-energy limit
is to recognize that the low-energy suppression of
scattering amplitudes (as well as the exact massless of
the light particle) is a consequence of the theory's
$U(1)$ symmetry. (The massless state has these
properties because it is this symmetry's
Nambu-Goldstone boson.\cite{ChiPT,GBreviews}) 
The simplicity of the low-energy
behaviour is therefore best displayed by:

\begin{enumerate}
\item
Making the symmetry explicit for the low-energy
degrees of freedom;
\item
Performing the low-energy approximation as early as
possible.
\end{enumerate}

\subsection{Exhibiting the Symmetry}

The $U(1)$ symmetry can be made to act exclusively
on the field which represents the light particle
by parameterizing the theory using a different set of variables
than ${\cal I}$ and ${\cal R}$. 
To this end imagine instead using polar coordinates in field space:
\begin{equation}
\label{polcoords}
\phi(x) = \chi(x) \; e^{i \theta(x)} .
\end{equation}
In terms of $\theta$ and $\chi$ the action of the 
$U(1)$ symmetry is simply $\theta \to \theta + \omega$, 
and the model's Lagrangian becomes: 
\begin{equation}
\label{linpolcoords}
{\cal L} = - \partial_\mu \chi \partial^\mu \chi - \chi^2 \partial_\mu \theta
\partial^\mu \theta - V(\chi^2).
\end{equation}
The semiclassical spectrum of this theory is found by expanding
${\cal L}$ in powers of the canonically-normalized
fluctuations, $\chi' = \sqrt2 (\chi - v)$ and $\theta'
= \sqrt2 \, v \, \theta$, about the vacuum $\chi = v$, revealing
that $\chi'$ describes the mass-$m_{\sst R}$ particle while
$\theta'$ represents the massless particle. 

With the $U(1)$ symmetry realized purely on the massless
field, $\theta$, we may expect good things to happen if we
identify the low-energy dynamics. 

\subsection{Timely Performance the Low-Energy Approximation}

To properly exploit the symmetry of the low-energy limit we
integrate out all of the high-energy degrees of freedom as the
very first step, leaving the inclusion of the low-energy degrees
of freedom to last. This is done most efficiently by computing
the following low-energy effective (or, Wilson) action. 

One way to split degrees of freedom into `heavy' and `light'
categories is to classify all field modes in momentum
space as heavy if (in Euclidean signature) 
they satisfy $p^2 + m^2 > \Lambda^2$ where $m$ is the
corresponding particle mass and $\Lambda$ is an 
appropriately chosen cutoff.\footnote{Although this
definition in terms of cutoffs most simply illustrates
the conceptual points of interest here, in practical calculations
it is usually dimensional regularization which is more
useful. Because modes of all frequencies appear in
dimensionally-regularized theories, the connection between
the effective theory and the underlying model whose low-energy
behaviour it describes is more subtle. It is most usefully
obtained by defining the effective lagrangian to be the
lagrangian which reproduces the low-energy 
amplitudes computed with the underlying theory order-by-order
in powers of the ratio of scales, 
$r$.\cite{Burgess95,ETbooks,ETreviews}} 

 Light modes are then all of
those which are not heavy. The cutoff, $\Lambda$,
which defines the boundary between these two kinds
of modes is chosen to lie well below the high-energy
scale (\ie\ well below $m_{\sst R}$ in the toy model)
but is also chosen to lie well above the low-energy scale
of ultimate interest (like the centre-of-mass energies, 
$E$, of low-energy scattering amplitudes). Notice that
in the toy model the heavy degrees of freedom defined
by this split include all modes of the field $\chi'$, as well 
as the high-frequency components of the massless field $\theta'$.

If $h$ and $\ell$ schematically denote the fields which are,
respectively, heavy
or light in this characterization, then the influence of heavy
fields on light-particle scattering at low energies is completely
encoded in the following effective lagrangian:
\begin{equation}
\label{leffdef}
\exp\left[i \int d^4x \; {\cal L}_{\rm eff}(\ell) \right]
= \int {\cal D}h \; \exp\left[ i \int d^4x \;
{\cal L}(\ell,h) \right]. 
\end{equation}

Physical observables at low energies are now computed 
by performing the remaining path integral over the light
degrees of freedom. By virtue of its definition, each
configuration in the integration over light fields is weighted by a 
factor of $\exp\left[i \int d^4x \; {\cal L}_{\rm eff}(\ell) \right]$
implying that the effective lagrangian weights the low-energy
amplitudes in precisely the same way as the classical lagrangian
does for the integral over both heavy and light degrees of freedom.

\subsection{Implications for the Low-Energy Limit}

Now comes the main point. When applied to the toy model
the condition of symmetry and the restriction to the low-energy
limit together have strong implications for ${\cal L}_{\rm eff}(\theta)$.
Specifically:

\begin{enumerate}
\item
Invariance of ${\cal L}_{\rm eff}(\theta)$ under the symmetry 
$\theta \to \theta + \omega$ implies ${\cal L}_{\rm eff}$ can
depend on $\theta$ only through the invariant quantity 
$\partial_\mu \theta$.
\item
Interest in the low-energy limit permits the expansion of 
${\cal L}_{\rm eff}$ in powers of derivatives of $\theta$.
Because only low-energy functional integrals remain
to be performed, higher powers of $\partial_\mu\theta$
correspond in a calculable way to higher suppression
of observables by powers of $E/m_{\sst R}$.
\end{enumerate}

Combining these two observations leads to the following
form for ${\cal L}_{\rm eff}$:
\begin{eqnarray}
\label{Leffform}
{\cal L}_{\rm eff} &=& - v^2 \; \partial_\mu \theta \, \partial^\mu \theta
+ a \, (\partial_\mu \theta \, \partial^\mu \theta)^2 
+ {b \over m_{\sst R}^2} \; (\partial_\mu \theta \, \partial^\mu \theta)^3
\nonumber\\
&& \qquad\qquad\qquad\qquad
+ {c \over m_{\sst R}^2} \; (\partial_\mu \theta \, \partial^\mu \theta)
\partial_\lambda \partial^\lambda
(\partial_\nu \theta \, \partial^\nu \theta) + \cdots,
\end{eqnarray}
where the ellipses represent terms which involve more than
six derivatives, and so more than two inverse powers of 
$m_{\sst R}$. A straightforward calculation comfirms this form in 
perturbation theory, with the additional information
\begin{equation}
\label{treecoeffs}
a_{\rm pert} = {1 \over 4 \lambda^2} + O(\lambda^0),
\qquad b_{\rm pert} = - \; {1 \over 4 \lambda^2} + O(\lambda^0),
\qquad c_{\rm pert} = {1 \over 4 \lambda^2}  + O(\lambda^0).
\end{equation}

In this formulation it is clear that each additional factor of 
$\theta$ is always accompanied by a derivative, 
and so implies an additional power of $r$ in its contribution 
to all light-particle scattering amplitudes. Because 
eq.~\pref{Leffform} is derived assuming only general properties of the
low energy effective lagrangian, its consequences (such as the
suppression by $r^n$ of low-energy $n$-point amplitudes) are
insensitive of the details of underlying model. They apply, in 
particular, to all orders in $\lambda$.

Conversely, the details of the underlying physics only enter 
through specific predictions, such as eqs.~\pref{treecoeffs}, 
for the low-energy coefficients $a,b$ and $c$. Different
models having a $U(1)$ Goldstone boson in their low-energy
spectrum can differ in the low-energy self-interactions of this
particle only through the values they predict for these
coefficients. 

\section{Lessons Learned}

It is clear that the kind of discussion given for the
toy model can be performed equally well for any other 
system having two well-separated energy scales.
There are a number of features of this example which
also generalize to these other systems. It is the purpose
of this section to briefly list some of these features.

\subsection{Why are Effective Lagrangians not More Complicated?}

${\cal L}_{\rm eff}$ as computed in the toy model is
not a completely arbitrary functional of its argument, $\theta$.
For example, ${\cal L}_{\rm eff}$ is {\sl real} and not
complex, and it is {\sl local} in the sense that 
(to any finite order in $1/m_{\sst R}$) 
it consists of a finite sum of powers of the field $\theta$
and its derivatives, all evaluated at the same point. 

Why should this be so?
Both of these turn out to be general features (so long as 
only massive degrees of freedom are integrated out) which
are inherited from properties of the underlying physics at
higher energies. 

\begin{itemize}
\item[{\it (i)}]
{\it Reality:} The reality of ${\cal L}_{\rm eff}$ 
is a consequence of the unitarity of the underlying
theory, and the observation that the degrees of 
freedom which are integrated out to obtain ${\cal L}_{\rm eff}$ 
are excluded purely on the grounds of their energy. As a
result, if no heavy degrees of freedom appear as part of an
initial state, energy conservation precludes their being produced
by scattering and so appearing in the final state. 

Since ${\cal L}_{\rm eff}$ is constructed to reproduce this
time evolution of the full theory, it must be real in order to
give a hermitian Hamiltonian as is required by unitary time
evolution.\footnote{There can be circumstances for which 
energy is not the criterion used to define the effective 
theory, and for which ${\cal L}_{\rm eff}$ is not real. The
resulting failure of unitarity in the effective theory
reflects the possibility in these theories of having states in 
the effective theory converting into states that have 
been removed in its definition.}

\item[{\it (ii)}] 
{\it Locality:} 
The locality of ${\cal L}_{\rm eff}$ is also a consequence 
of excluding high-energy states in its definition,
together with the Heisenberg Uncertainty Relations. Although
energy and momentum conservation preclude the
direct production of heavy particles (like those described by 
$\chi$ in the toy model) from an initial low-energy particle
configuration, it does not preclude their {\it virtual} production.

That is, heavy particles may be produced so long as they are
then re-destroyed sufficiently quickly. Such virtual production
is possible because the Uncertainty Relations permit 
energy to be not precisely conserved for states which do
not live indefinitely long. A virtual state whose production
requires energy nonconservation of order $\Delta E \sim 
M$ therefore cannot live longer than $\Delta t \sim 1/M$,
and so its influence must appear as being local in time
when observed only with probes having much smaller
energy. Similar arguments imply locality in space for 
momentum-conserving systems. 

Since it is the mass $M$ of the heavy particle which sets
the scale over which locality applies once it is integrated
out, it is $1/M$ which appears with derivatives of 
low-energy fields when ${\cal L}_{\rm eff}$ is written
in a derivative expansion.

\end{itemize}

\subsection{Predictiveness and Power Counting}

The entire rationale of an effective lagrangian is
to incorporate the virtual effects of high-energy particles
in low-energy processes, order-by-order in powers of
the small ratio, $r$, of these two scales ({\it e.g.}
$r = E/m_{\sst R}$ in the toy model). In order to 
use an effective lagrangian it is therefore necessary
to know which terms contribute to physical processes
to any given order in $r$. 

This determination is explicitly possible if the low-energy degrees 
of freedom are weakly interacting, because in this case perturbation
theory in the weak interactions may be analyzed graphically,
permitting the use of powercounting arguments to
systematically determination where powers
of $r$ originate. Notice that the assumption of a weakly-interacting
low-energy theory does {\it not} presuppose the underlying
physics to be also weakly interacting. For instance, for the
toy model the Goldstone boson of the low-energy theory is
weakly interacting provided only that the $U(1)$ symmetry 
is spontaneously broken, and this is true 
{\it independent} of the size of $\lambda$. 

For example, in the toy model the effective lagrangian takes the
general form: 
\begin{equation}
\label{genformLeff}
{\cal L}_{\rm eff} = v^2 m_{\sst R}^2 \sum_{id} {c_{id} 
\over m_{\sst R}^d} \; {\cal O}_{id},
\end{equation}
where the sum is over interactions, ${\cal O}_{id}$, 
involving $i$ powers of the field $\theta$ and $d$ 
derivatives. The power of $m_{\sst R}$ premultiplying
each term is chosen to ensure that the coefficient 
$c_{id}$ is dimensionless. (For instance, the interaction
$(\partial_\mu \theta \, \partial^\mu \theta)^2$ has
$i = d = 4$.) There are two useful properties which
all of the operators in this sum must satisfy:
\begin{enumerate}
\item
$d$ must be even by virtue of Lorentz invariance.
\item
Since the sum is only over interactions,
it does not include the kinetic term, which is the unique
term for which $d=i=2$.
\item
The $U(1)$ symmetry implies every factor of $\theta$
is differentiated at least once, and so $d \ge i$. Furthermore,
any term linear in $\theta$ must therefore be a total 
derivative, and so may be omitted, implying 
$i \ge 2$ without loss. 
\end{enumerate}

It is straightforward to powercount \cite{Burgess95} the powers
of $v$ and $m_{\sst R}$ that interactions of this form 
contribute to an $\ell$-loop contribution to
$n$-point Goldstone-boson scattering 
amplitude, ${\cal A}_{n\ell}(E)$, at centre-of-mass energy $E$:
\begin{eqnarray}
\label{PCresult1}
{\cal A}_{n\ell}(E) &\sim& v^2 m_{\sst R}^2  \left( {1 \over v} 
\right)^n \left( { m_{\sst R} \over 4 \pi v} \right)^{2\ell}
\left( {E \over m_{\sst R}} \right)^P, \\
\hbox{where} \qquad P &=& 2 + 2\ell + \sum_{id} (d-2) V_{id}.
\label{PCresult2}
\end{eqnarray}
Here $V_{id}$ counts the number of times an interaction involving
$i$ powers of fields and $d$ derivatives appears
in the amplitude. Eqs.~\pref{PCresult1} and \pref{PCresult2}
have several noteworthy features:
\begin{enumerate}
\item
The factor $(m_{\sst R}/4\pi v)^{2\ell}$ can ruin
the perturbative expansion if $m_{\sst R}$ were to
be too much larger than $v$. Since, in the toy model,
$m_{\sst R} = \lambda v$, this factor is simply
of order $(\lambda/4 \pi )^{2 \ell}$, which reproduces the
coupling-constant dependence of loops in the underlying theory.
\item
The condition $d \ge i \ge 2$, and the omission of
the case $d=i=2$, ensures that all of the
terms in the expression for $P$ are positive. All graphs
are therefore suppressed by some power of $r 
= E/m_{\sst R}$. Furthermore, it is straightforward
to identify the graphs which contribute to ${\cal A}_{n\ell}$
to any fixed order in $r$. 
\end{enumerate}

To see how eqs.~\pref{PCresult1} and \pref{PCresult2}
are used, consider the first few orders of
$r$ in the toy model. $P=4$ is the smallest value possible (since $d$
must be even), and arises only if $\ell = 0$ and if $V_{i4}=1$, 
all others zero (for a single $d=4$ vertex).
Because $i \le d$, an $O(r^4)$ contribution can therefore
arise only for $n  \le 4$. 

The utility of powercounting really becomes clear when 
subleading behaviour is computed. $P=6$ is
achieved if and only if either: ({\it i}) $\ell = 1$ and $V_{i4} = 1$,
with all others zero; or ({\it ii}) $\ell = 0$ and $\sum_{i}
\Bigl(4 V_{i6} + 2 V_{i4} \Bigr) = 4$. The only choice
which combines into a 4-point amplitude ($n=4$) is
therefore a tree graph ($\ell = 0$) involving two $d=4$
3-point vertices, $V_{34} = 2$. 

\subsection{The Effective Lagrangian Logic}

With the powercounting results in hand we can see how
to calculate predictively --- {\it including loops} --- using 
the nonrenormalizable effective theory. The logic follows
these steps:

\begin{enumerate}
\item %1
Choose the accuracy desired in the answer. (For instance
an accuracy of 1\% might be desired in a particular scattering
amplitude.)
\item %2
Determine the order in the small ratio of scales (\ie\
$r = E/m_{\sst R}$ in the toy model) which
is required in order to achieve the desired accuracy. (For
instance if $r = 0.1$ then $O(r^2)$ is required to achieve
1\% accuracy.)
\item %3
Use the powercounting results to identify which terms in
${\cal L}_{\rm eff}$ can contribute to the observable
of interest to the desired order in $r$. At any fixed order
in $r$ this always requires a finite number (say: $N$) of terms in
${\cal L}_{\rm eff}$.
\item[4a.] %4a
If the underlying theory is known, and is calculable, then
compute the required coefficients of the $N$ required
effective interactions to the accuracy required. (In the
toy model this corresponds to calculating the coefficients
$a,b,c$ {\it etc.}
\item[4b.] %4b
If the underlying theory is unknown, or is too complicated
to permit the calculation of ${\cal L}_{\rm eff}$, then
leave the $N$ required coefficients as free parameters.
The procedure is nevertheless predictive if more than
$N$ observables can be identified whose predictions 
depend only on these parameters. 

\end{enumerate}

Step 4a is required when the low-energy expansion
is being used as an efficient means to accurately
calculating observables in a well-understood theory.
It is the option of choosing instead
Step 4b, however, which introduces much of
the versatility of effective-lagrangian methods. Step
4b is useful both when the underlying theory is not known
(such as when searching for physics beyond the 
Standard Model) {\it and} when the underlying physics
is known but complicated (like when describing the
low-energy interactions of pions in Quantum Chromodynamics).

The effective lagrangian is in this way seen to be predictive
even though it is not renormalizable in the usual sense. In
fact, renormalizable theories are simply the special case
of Step 4b where one stops at order $r^0$, and so are the
ones which dominate in the limit that the light and heavy
scales are very widely separated. 
We see in this way {\it why} renormalizable interactions 
play ubiquitous roles through physics!
These observations have important
conceptual implications for the quantum behaviour of
other nonrenormalizable theories, such as gravity, to which
we return in the next section. 

\subsection{The Choice of Variables}

The effective lagrangian of the toy model seems to
carry much more information when $\theta$ is used to 
represent the light particles than it would if ${\cal I}$ 
were used.
How can physics depend on the fields which are used
to parameterize the theory?

Physical quantities do not depend on what variables are
used to describe them, and the low-energy
scattering amplitude is suppressed by the same power
of $r$ in the toy model regardless of whether it is
the effective lagrangian for ${\cal I}$ or $\theta$
which is used at an intermediate stage of the calculation.

The final result would nevertheless
appear quite mysterious if ${\cal I}$ were
used as the low-energy variable, since it would emerge as
a cancellation only at the end of the calculation. With
$\theta$ the result is instead manifest at every step. 
Although the physics does not depend on the variables
in terms of which it is expressed, it nevertheless pays
mortal physicists to use those variables which make
manifest the symmetries of the underlying system.

\subsection{Regularization Dependence}

The definition of ${\cal L}_{\rm eff}$ appears to
depend on lots of calculational details, like the value of $\Lambda$
(or, in dimensional regularization, the matching scale) and
the minutae of how the cutoff is implemented. Why doesn't
${\cal L}_{\rm eff}$ depend on all of these details? 

${\cal L}_{\rm eff}$ generally {\sl does} 
depend on all of the regularizational details. But these
details all must cancel in final expressions for physical
quantities. Thus, some $\Lambda$-dependence enters into
scattering amplitudes through the explicit dependence which 
is carried by the couplings of ${\cal L}_{\rm eff}$ (beyond
tree level). But $\Lambda$ also potentially enters scattering
amplitudes because loops over all light degrees of freedom
must be cut off at $\Lambda$ in the effective theory, by
definition. The cancellation of these two sources of
cutoff-dependence is guaranteed by the observation that
$\Lambda$ enters only as a bookmark, keeping
track of the light and heavy degrees of freedom at
intermediate steps of the calculation. 

This cancellation of $\Lambda$ in all physical quantities
ensures that we are free to make any choice of cutoff 
which makes the calculation convenient. After all, although
all regularization schemes for ${\cal L}_{\rm eff}$ give
the same answers, more work is required for 
some schemes than for others. Mere mortal physicists
use an inconvenient scheme at their own peril!

This freedom to use {\sl any} convenient
scheme is ultimately the reason why dimensional regularization 
may be used when defining low-energy effective theories,
even though the dimensionally-regularized effective
theories involve fields with modes of arbitrarily high
momentum. So long as the effective interactions are
chosen to properly reproduce the dimensionally-regularized 
scattering amplitudes of the full theory (order-by-order
in $1/M$) any regularization-dependent properties will
necessarily drop out of the final results. 

\subsection{The Meaning of Renormalizability}

The previous discussion about the cancellation between
the cutoffs on virtual light-particle momenta and
the explicit cutoff-dependence of ${\cal L}_{\rm eff}$
is eerily familiar. It echoes the traditional discussion
of the cancellation of the regularized ultraviolet divergences of
loop integrals against the regularization dependence of
the counterterms of the renormalized lagrangian. There
are, however, the following important differences.

\begin{enumerate}
\item
The cancellations in the effective theory occur even though
$\Lambda$ is not sent to infinity, and even though 
${\cal L}_{\rm eff}$ contains arbitrarily many terms which
are not renormalizable in the traditional sense (\ie\ terms
whose coupling constants have dimensions of inverse powers
of mass in fundamental units where $\hbar =  c = 1$). 
\item
Whereas the cancellation of regularization dependence
in the traditional renormalization picture appears {\it ad-hoc}
and implausible, those in the effective lagrangian are sweet
reason personified. This is because they simply
express the obvious fact that $\Lambda$ only was introduced
as an intermediate step in a calculation, and so {\sl cannot}
survive uncancelled in the answer. 
\end{enumerate}

This resemblance suggests Wilson's physical reinterpretation of
the renormalization procedure. Rather than considering a model's
classical lagrangian, such as ${\cal L}$ of eq.~\pref{abeltoymodel}, 
as something pristine and fundamental, it is
better to think of it also as an effective lagrangian obtained
by integrating out still more microscopic degrees of freedom. 
The cancellation of the ultraviolet divergences in this interpretation
is simply the usual removal of an intermediate step in an
calculation to whose microscopic part we are not privy.

\section{Quantum Gravity: A Conceptual Payoff}

According to the approach just described, nonrenormalizable
theories are not fundamentally different from renormalizable
ones. They simply differ in their sensitivity to more microscopic
scales which have been integrated out. It is instructive to 
see what this implies for the nonrenormalizable theories which
sometimes are required to successfully describe experiments.
This is particularly true for the most famous such case, 
Einstein's theory of gravity. 

\subsection{The Effective Theory of Gravity}

The low-energy degrees of freedom in this case are the
metric, $g_{\mu\nu}$, of spacetime itself. As has been seen
in previous sections, Einstein's action for this theory should 
be considered to be just one term in a sum of all possible 
interactions which are consistent with the symmetries of
the low-energy theory (which in this case are:
general covariance and local Lorentz invariance):\footnote{A
cosmological constant is not written here, because the observed
size of the universe implies this is extremely small. There is no
theoretical understanding why the cosmological constant should
be so small.}
\begin{equation}
\label{gravaction}
{\cal L}_{\rm eff} = \sqrt{- g} \left( \hf \, M_p^2 \, R
+ a \, R_{\mu\nu} \, R^{\mu\nu} + b \, R^2 + {c \over m^2} \; R^3
+ \cdots \right) .
\end{equation}
Here $R_{\mu\nu}$ is the metric's Ricci tensor, $R$ is the
Ricci scalar, and a term involving $R_{\mu\nu\lambda\rho}
R^{\mu\nu\lambda\rho}$ is not written because (in four
dimensions) it can be rewritten in terms of those displayed,
plus a total derivative. All we need to know about these 
quantities is that they each involve two derivatives 
of the field $g_{\mu\nu}$. The term linear in $R$ is the usual
Einstein action of General Relativity.
Only one representative of the many
possible curvature-cubed terms is explicitly written. 
The constants appearing in ${\cal L}_{\rm eff}$
are the Planck mass $M_p \sim 10^{18}$ GeV and
several dimensionless constants, $a,b$ and $c$.

The mass scale $m$ can be considered as the
smallest microscopic scale to have been integrated out to obtain
eq.~\pref{gravaction}. For definiteness we might take
the electron mass, $m = 5\times 10^{-4}$ GeV, for $m$ when
considering applications at energies below the masses of
all elementary particles. (Notice that contributions like
$m^2 R$ or $R^3/M_p^2$ could also exist,
but these are completely negligible compared to the
terms displayed in eq.~\pref{gravaction}.)

\subsection{Powercounting}

Since gravitons are weakly coupled, perturbative powercounting
may be used to see how the high-energy scales $M_p$ and $m$
enter into observables like graviton scattering amplitudes about
some fixed macroscopic metric (like flat space). 

The $\ell$-loop contribution to the $n$-point amplitude which
involves $V_{id}$ vertices involving $d$ derivatives and
the emission or absorption of $i$ gravitons turns
out to be of order:
\begin{eqnarray}
\label{GRcount1}
{\cal A}_{n\ell}(E) &\sim& m^2 M_p^2 \left( {1 \over M_p}
\right)^n \left( {m \over 4 \pi M_p} \right)^{2 \ell} 
\left( {m^2 \over M_p^2} \right)^{\sum_{id}' V_{id}}
\left( {E \over m} \right)^P \nonumber\\
\hbox{where} \qquad P &=& 2 + 2L + \sum_{id}
(d-2) V_{id}.
\label{GRcount2}
\end{eqnarray}
The prime on the sum in the exponent of the penultimate
terms indicates the omission of the case $d=2$ from 
the sum over $d$. 

Eqs.~\pref{GRcount1} and \pref{GRcount2} share
the many noteworthy features of eqs.~\pref{PCresult1}
and \pref{PCresult2}. There are some features that are
peculiar to gravity, however:

\begin{enumerate}
\item
Unlike for the toy model the term involving
two derivatives includes interactions as well as the kinetic
terms, and so $d=2$ is included in the sum which appears
in the definition of $P$. 
\item
There is, in addition to the factor $(m/4 \pi M_p)^{2\ell}$,
a further suppression by powers of $m/M_p$ every time
a vertex taken any term in ${\cal L}_{\rm eff}$ with
the exception of the Einstein term. But the relative
suppression of $d = 8$ terms relative to $d=6$ terms
comes purely from powers of $E/m$ rather than
$m/M_p$ (for any fixed number of loops). 
\end{enumerate}

The explicit expression for $P$ permits a determination
of the dominant low-energy contributions to scattering
amplitudes. The minimum suppression comes when
$\ell = 0$ and $P = 2$, and so is given by arbitrary
tree graphs constructed purely from the Einstein action.
We are led in this way to what we in any case believe:
it is classical General Relativity which governs the 
low-energy dynamics of gravitational waves!

But the next-to-leading contributions are also quite
interesting. These arise in one of two ways, either:
({\it i}) $\ell = 1$ and $V_{id} = 0$ for any $d\ne 2$;
or ({\it ii}) $\ell = 0$, $\sum_i V_{i4} = 1$, 
$V_{i2}$ is arbitrary, and all other $V_{id}$ vanish.
That is, the next to leading contribution is obtained by
computing the one-loop corrections using only Einstein
gravity, or by working to tree level and including precisely
one curvature-squared interaction in addition to any
number of interactions from the Einstein term. Both
are suppressed compared to the leading term by a factor
of $(m/M_p)^2(E/m)^2 = (E/M_p)^2$, and the one-loop
contribution carries an additional factor of $(1/4\pi)^2$. 

\subsection{Predictability}

Working to next-to-leading order, the effective lagrangian
contains three unknown parameters, $M_p$, $a$ and $b$.
Using option 4b of the Effective Lagrangian Logic permits
these to be fit from gravitational experiments, while
still giving real predictions, so long as
at least four observables are considered. 

In fact, next-to-leading calculations along these
lines were made in ref.~\cite{DonoghueGR}, 
where it is also pointed out how to distinguish the
quantum contributions from those arising from
the curvature-squared interactions. They may be
distinguished (in principle) from the dependence
of the gravitational potential energy of two masses
on the separation between the masses. Reinstating
powers of $\hbar$ and $c$, the potential energy 
including next-to-leading corrections may be written:
\begin{equation}
\label{GRcorr}
V(r) = - \; {G M_1 M_2 \over r} \left\{ \left[ 1 - A \; 
{G(M_1 + M_2) \over r c^2} + \cdots \right]
+ B \; {G \hbar \over r^2 c^3} + \cdots \right\},
\end{equation}
where $G = 1/(16 \pi M_p^2)$ is Newton's constant,
$M_1$ and $M_2$ are the masses whose potential 
energy is of interest, and which are separated by
a distance $r$.
The square brackets, $\Bigl[1 + \cdots \Bigr]$, in this
expression represent the relativistic corrections to the 
Newtonian potential which already arise within classical
General Relativity, and which must all be included
in the leading-order calculation. ($A$ here
is a known constant whose
numerical value plays no role in what follows.) 

It is the term proportional to $\hbar$ which expresses
the one-loop contribution. The main point is that this 
contribution cannot be confused with any of the others,
because the curvature-squared terms do not contribute
to any finite order in $1/r$, and the classical relativistic
terms depend differently on $G$ and $M_k$. The
coefficient $B$ is finite and is an absolute prediction
of quantum General Relativity. (Ref.~\cite{DonoghueGR}
computes most, but not all, of the graphs which 
contribute to $B$.)

\subsection{Quantum Gravity}

Table 1 shows that the quantum-gravitational correction just 
discussed is numerically small when evaluated for 
garden-variety gravitational fields. 

\vspace{3mm}
\begin{center}
\begin{tabular}{lcc}
& $r = R_\odot$ & $r = 2 GM_\odot/c^2$ \\
&&\\
$GM_\odot/rc^2$ & $10^{-6}$ & $0.5$ \\
$G\hbar/r^2c^3$ & $10^{-88}$ & $10^{-76}$ 
\end{tabular}
\end{center}

\begin{quote}
{\footnotesize               
{\bf Table 1:} {\sl The size of relativistic and quantum corrections
to the Sun's gravitational field.}}
\end{quote}

Of course, the point of these numbers is not to argue
that any such quantum corrections are likely to be detected
in the forseeable future. Rather, the small size of these
quantum corrections instead show that the experimental 
great success of classical General Relativity in the
solar system can also be considered a great success
of {\it quantum} gravity! Classical calculations are 
not a poor substitute for some poorly-understood
quantum theory, they are rather an extremely good
approximation for which quantum corrections are 
exceedingly small. 

This is also not to say that the vexing problem
of quantum gravity is in any way solved. Among the
deep unsolved issues are understanding the quantum
nature of spacetime and the gravitational behaviour of
systems under extreme conditions. What the
effective-lagrangian perspective gives is a focussing of
issues. By showing where quantum gravity is under complete
control --- \ie\ for long-distance, macroscopic fields such
as arise in the solar system --- effective lagrangian methods
direct attention to where the burning issues really lie. 

What better way to exemplify the beauty and power 
of effective lagrangian
techniques on both the practical and conceptual levels?

\section*{Acknowledgements}
I am indebted to Prof. Joan Sola and the organizers of
RADCOR 98 for the kind invitation to give these lectures, as 
well as their warm hospitality while I was in Barcelona.

These lectures are partially based on a longer series of lectures
given in E.T.H. Lausanne and the Universit\'e de
Neuch\^atel in June 1995, entitled 
{\it An Introduction to Effective Lagrangians and Their Applications}. 

My principal research funds come from {\sl the Natural Sciences and 
Engineering Council of Canada}, with some additional funds 
being provided by {\sl les Fonds pour la Formation de Chercheurs 
et l'Aide \`a la R\'echerche du Qu\'ebec}. 

%\section*{References}

\end{document}